\documentclass[a4paper,11pt]{article}
\usepackage{jheppub}
\usepackage{hyperref}
\usepackage{epstopdf,amsfonts,appendix,comment}
\usepackage{color,slashed,footnote,multirow,longtable,braket}
\usepackage{mathrsfs,latexsym,color,url,etoolbox}
\usepackage{physics}
\usepackage[normalem]{ulem}
\usepackage[table]{xcolor}
\usepackage[capitalize]{cleveref}
\usepackage{floatrow}
\usepackage{graphicx}
\usepackage{subcaption}
\newfloatcommand{capbtabbox}{table}[][\FBwidth]

\graphicspath{{figures/}}

\title{\boldmath Beyond $\Delta m^2$: Absolute Mass Sensitivity in Neutrino Oscillations}

\author[1,2]{Gustavo F.~S. Alves,}
\author[2]{André L.~C.~de Gouvêa,}
\author[3]{Joshua Kaler,}
\author[3]{Shirley Weishi Li,}
\author[1]{Pedro A.~N.~Machado}
\affiliation[1]{Theoretical Physics Department, Fermilab, P.O. Box 500, Batavia, IL 60510, USA}
\affiliation[2]{Northwestern University, Department of Physics \& Astronomy, 2145 Sheridan Road, Evanston, IL 60208, USA}
\affiliation[3]{Department of Physics and Astronomy, University of California, Irvine, CA 92697, USA}
\emailAdd{gustavo.alves@northwestern.edu}
\emailAdd{degouvea@northwestern.edu}
\emailAdd{kalerj@uci.edu}
\emailAdd{shirley.li@uci.edu}
\emailAdd{pmachado@fnal.gov}

\dedicated{Dedicated to the memory of Boris Kayser, who is partially to blame for this.}

\date{August 13, 2026}

\abstract{
Conventional wisdom says that neutrino oscillations measure only mass-squared differences and not the absolute neutrino mass scale. 
This is true, however, only at leading order in the expansion parameters $m_i/E$, the ratios of the neutrino masses $m_i$ ($i=1,2,3$) to the neutrino energy $E$.
At next-to-leading order, the oscillation phase includes terms proportional to $m_i^4-m_j^4 = \Delta m^2_{ij}(m_i^2+m_j^2)$, and is therefore sensitive to the absolute mass scale. 
In this paper, we derive the next-to-leading-order corrections using a wave-packet treatment and taking into account the neutrino-production kinematics. 
We then apply this result to reactor antineutrinos and find that the JUNO experiment is sensitive to neutrino masses of a few hundred keV. 
While not competitive with existing bounds from beta decay, electron capture, and cosmology, neutrino oscillations provide a novel, complementary probe of the neutrino mass scale, with sensitivity to a different combination of neutrino masses.
}

\begin{document}

\preprint{FERMILAB-PUB-26-0584-T, UCI-HEP-TR-2026-11}
\maketitle
\flushbottom


\section{Introduction}
\label{sec:intro}

Neutrino oscillations arise from the misalignment between the flavor eigenstates $\nu_\alpha$ ($\alpha = e,\mu,\tau$) and the mass eigenstates $\nu_i$ ($i = 1,2,3$), with the two bases related by the leptonic mixing matrix $U$: $\nu_\alpha = \sum_i U_{\alpha i}^*\,\nu_i$. As neutrinos propagate in spacetime, different mass eigenstates accumulate different phases, leading to flavor oscillations. In the standard ultrarelativistic treatment of neutrino propagation and flavor evolution, the phase difference is $\Delta m^2_{ij}L/2E$, where $\Delta m^2_{ij}\equiv  m_i^2-m_j^2$, $L$ is the baseline, and $E$ is the neutrino energy. This is the origin of the conventional wisdom that oscillation experiments are insensitive to the absolute neutrino mass scale: They depend only on mass-squared differences.

The standard result, however, is approximate: It keeps only the leading-order (LO) term in the expansion of the propagation phase in powers of $m_i/E$.
At higher order, new mass combinations can appear in the oscillation phase.
In this paper, we show that the next-to-leading-order (NLO) correction to the oscillation phase is proportional to $(m_i^4-m_j^4)L/E^3$. 
Because this term depends on the product of $m_i^2+m_j^2$ and $m_i^2-m_j^2$, neutrino oscillations are in principle sensitive to the absolute neutrino mass scale.

The computation of the NLO correction to the oscillation phase is less straightforward than that of the LO result.
At LO, different simplifying assumptions, such as using the ``equal-energy'' or ``equal-momentum'' prescriptions, and assuming the naive speed-of-light relation between the propagation time and the baseline ($t=L$), lead to the same (correct) oscillation phase. At NLO, these assumptions are no longer equivalent, nor do they lead to the correct result. 
A proper treatment requires keeping track of the wave-packet nature of the neutrino state and carefully considering the kinematics of the neutrino production process.

The Jiangmen Underground Neutrino Observatory (JUNO) provides a useful benchmark for estimating the impact of the NLO correction on the oscillation phase and the resulting sensitivity to the absolute neutrino mass scale. Its long baseline, excellent energy resolution, and large expected statistics make it especially sensitive to different features of the oscillation phase, as already demonstrated by its world-leading measurements of the oscillation parameters $\sin^2\theta_{12}$ and $\Delta m^2_{21}$~\cite{JUNO:2025gmd}. 
We use JUNO as a realistic case study to assess how well neutrino oscillations could constrain the absolute mass scale. While the NLO effect is too small to translate into sensitivity to the absolute neutrino masses that is competitive with the most stringent bounds on the neutrino mass scale, we believe that neutrino oscillations serve as a novel and complementary probe. The effect directly impacts the oscillation phase and is sensitive to a different combination of the neutrino masses. 

The rest of this paper is organized as follows. 
In Sec.~\ref{sec:beyond-leading-order}, we derive the next-to-leading-order oscillation phase using a wave-packet treatment and including the necessary details of the neutrino production kinematics. 
In Sec.~\ref{sec:JUNO_study}, we apply our result to JUNO and estimate its sensitivity to the absolute neutrino mass scale. In Sec.~\ref{sec:mass_limits}, we summarize existing absolute mass constraints and compare them with the projected JUNO sensitivity. We present our conclusions in Sec.~\ref{sec:conclusions}.


\section{Neutrino Oscillations Beyond Leading Order}
\label{sec:beyond-leading-order}

There is a ubiquitous heuristic description of neutrino oscillations that can be summarized as follows. At production, the neutrino is described as a linear superposition of the different neutrino mass eigenstates. These are treated as plane waves that satisfy the standard dispersion relation for massive particles, $E^2=|\vec{p}|^2+m_i^2$. One then assumes the different components of the neutrino state to have either a common energy or a common linear momentum, and the propagation time is related to the baseline by taking $t=L$. With these ingredients, it is straightforward to compute the phase differences among the propagating mass eigenstates. None of these assumptions are physically justified, as has been extensively discussed in the literature~(see, for example, Refs.~\cite{Nussinov:1976uw,Kayser:1981ye,Giunti:1991ca,Grimus:1996av,Giunti:1997wq,Grimus:1998uh,Beuthe:2001rc,Akhmedov:2009rb,Akhmedov:2010ms,Akhmedov:2019iyt,Giunti:2002xg,Giunti:2003ax,Shirokov:2006yf,Grimus:2019hlq,Naumov:2020yyv}). Nonetheless, assuming the neutrinos are ultrarelativistic, $m_i\ll E,|\vec{p}|$, different ad hoc assumptions yield the same oscillation phase at leading order (LO) in $m_i/E$, and all agree with the LO results of a more careful treatment of the phenomenon.

Beyond LO, this is no longer true. 
The equal-energy and equal-momentum prescriptions, for example, lead to different answers at next-to-leading order (NLO) in $m_i/E$, as we discuss in more detail in Appendix~\ref{app:NLOplanewaves}.
A more careful description of the neutrino state and how it evolves in spacetime is required in order to compute the NLO correction to the oscillation phase. 
Such a description needs to unambiguously relate the propagation time to the detection position, determine the mean energy and momentum of each mass eigenstate from the production process, and specify the many different relevant approximations clearly. We do this using the wave-packet formalism for neutrino propagation~\cite{Nussinov:1976uw, Giunti:1991ca, Giunti:1997wq, Beuthe:2001rc, Akhmedov:2010ms, Akhmedov:2019iyt, Giunti:2002xg, Shirokov:2006yf, Bernardini:2006ak} and by properly investigating the kinematical constraints imposed by the neutrino production process. Since only the propagation direction is relevant for oscillation probabilities, we use a one-dimensional treatment throughout the paper. Our results can be trivially extended to three space dimensions. 
 

\subsection{Wave packets and the oscillation probability density}
\label{subsec:wave-packet}

A neutrino produced as a flavor eigenstate can be described as a coherent superposition of mass-eigenstate wave packets. Each mass eigenstate with mass $m_i$ will be described as a Gaussian in momentum space, sharply peaked at mean momentum $P_i$ and with a common width $\sigma_p$. In general, the different mass eigenstates have slightly different widths, but, as we will see below, $\sigma_p$ does not enter our final result and the common-width assumption is inconsequential. Assuming that a $\nu_\alpha$ is produced at $t=0$ and localized at $x=0$, its time-evolved wave function is
\begin{equation}
    |\nu_\alpha(x,t)\rangle= \sum_i \frac{U_{\alpha i}^*}{\sqrt{2\pi}} \int 
    \frac{dp}{(2\pi\sigma_p^2)^{1/4}}\exp\left(-\frac{(p - P_i)^2}{4\sigma_p^2} + ipx - iE_i(p)t\right)  |\nu_i\rangle,
\label{eq:wave-packet}
\end{equation}
where $E_i(p) = \sqrt{p^2 + m_i^2}$. 

To perform the integral analytically, we expand each energy $E_i(p)$ around its mean momentum $P_i$,
\begin{equation}
    E_i(p) = E_i(P_i) + v_i(p - P_i) + \mathcal{O}\left(\frac{m_i^2(p-P_i)^2}{E_i(P_i)^3}\right),
\end{equation}
with group velocity $v_i = P_i/E_i(P_i)$, and keep terms through linear order in $p-P_i$. Because the wave packet restricts $|p-P_i|\lesssim\sigma_p$, this is a narrow wave packet expansion in $\sigma_p / P_i$ and is distinct from the relativistic expansion in $m_i/E$ we perform later.

With this approximation, the integral in Eq.~(\ref{eq:wave-packet}) is Gaussian and can be solved analytically:
\begin{equation} 
\label{eq:wavepacket}
    |\nu_\alpha(x,t)\rangle = \sum_i \frac{U_{\alpha i}^*}{(2\pi\sigma_x^2)^{1/4}}\,
    \exp\!\left[-\frac{(x - v_i t)^2}{4\sigma_x^2} + i P_i x - iE_i(P_i) t \right]|\nu_i\rangle ,
\end{equation}
where $\sigma_x \equiv 1/(2\sigma_p)$ is the spatial width of the wave packet. Equation~\eqref{eq:wavepacket} shows that $\sigma_x$ controls how strongly $x$ and $t$ are correlated: for a spatially-localized wave packet, the wave function is strongly peaked near the classical trajectory $x = v_i t$.

The propagated neutrino is detected via the weak interactions, which cannot distinguish among the mass eigenstates and project the neutrino onto a flavor eigenstate $\nu_\beta$, with $\langle\nu_\beta|\nu_i\rangle=U_{\beta i}$. Hence, the amplitude to detect the neutrino as flavor $\beta$ is $A_{\alpha \beta} = \langle \nu_\beta | \nu_\alpha (x, t) \rangle$ and the probability density is $|A_{\alpha\beta}|^2$. Using Eq.~\eqref{eq:wavepacket},
\begin{align}
    \begin{split}
       & |A_{\alpha\beta}(x, t)|^2
       =\sum_{i,j} U_{\alpha i}^*\, U_{\beta i}\, U_{\alpha j}\, U_{\beta j}^*\\
       &\qquad \quad \times\frac{1}{(2\pi\sigma_x^2)^{1/2}}\,
        \exp\!\left[-\frac{(x - v_i t)^2 + (x - v_j t)^2}{4\sigma_x^2} +i\Delta P_{ij}\,x -i\Delta E_{ij}t\right],
    \end{split}
    \label{eq:prob_density}
\end{align}
with $\Delta P_{ij} = P_i - P_j$ and $\Delta E_{ij} = E_i(P_i) - E_j(P_j)$. 

The next steps are as follows. First, we need to relate the probability density $|A_{\alpha\beta}(x,t)|^2$ (probability per unit length) to a probability. 
Second, $|A_{\alpha\beta}(x,t)|^2$ depends on both the position $x$ and the time $t$, whereas the oscillation probability extracted from neutrino oscillation data is a function of the baseline $L$ alone. We construct the oscillation probability measured by experiments carefully in the next subsection.


\subsection{The oscillation probability measured by an experiment}
\label{subsec:measured-prob}

We want to relate Eq.~\eqref{eq:prob_density} to the probability inferred by an experiment, $P_{\alpha\beta}(L)$. This requires three steps: integrating over the finite detection region, accounting for the unmeasured propagation time, and normalizing to a conditional probability.

We first compute the probability that the neutrino is detected inside a finite region within the detector, located around the baseline position $L$. This requires integrating $|A_{\alpha\beta}(x, t)|^2$ over the detection region. 
To keep the integrals and the answers analytic, we approximate the detection region by a Gaussian centered at $L$ with width $\delta L$, and extend the integral range to infinity. We define the ancillary quantity
\begin{equation} 
    \widetilde{P}_{\alpha\beta}(L,t) = \int_{-\infty}^{\infty} dx\; e^{-\frac{(x-L)^2}{2\delta L^2}}\;|A_{\alpha\beta}(x,t)|^2\, .
\label{eq:unnorm_prob} 
\end{equation}
Because the Gaussian window is dimensionless while $|A_{\alpha\beta}|^2$ carries dimensions of inverse length, $\widetilde{P}_{\alpha\beta}(L,t)$ is dimensionless. It is proportional to the joint probability that the neutrino is detected in the detection region at time $t$ with flavor $\beta$. 

Two things still separate $\widetilde{P}_{\alpha\beta}(L,t)$ from $P_{\alpha\beta}(L)$.
First, $P_{\alpha\beta}(L)$ is normalized such that $\sum_\beta P_{\alpha\beta}(L)=1$ for all $\alpha$ and $L$. Instead, $\sum_\beta \widetilde{P}_{\alpha\beta}(L,t)$ is the probability of detecting a neutrino at time $t$, with any flavor, in the detection region (as opposed to somewhere else). This is, of course, less than one for a finite detector. Second, $\widetilde{P}_{\alpha\beta}(L,t)$ still depends on the propagation time $t$, whereas $P_{\alpha\beta}(L)$ does not. We first resolve the time dependence by examining how experiments infer $P_{\alpha\beta}(L)$; we address the normalization issue in the conditional-probability step that follows.

The prototypical oscillation experiment consists of a source located at the origin characterized by a neutrino flux $\Phi^S_{\alpha}(t)$ for each neutrino flavor $\nu_{\alpha}$ ($\alpha=e,\mu,\tau$), together with a detector a distance $L$ away. The flux of $\nu_{\beta}$ ($\beta=e,\mu,\tau$) at the detector is defined to be\footnote{In three dimensions, the flux at the detector site would be rescaled by a factor proportional to $L^{-2}$.}
\begin{equation}
\Phi_{\beta}^D(t) = \sum_\alpha P_{\alpha\beta}(L)\Phi^S_{\alpha}(t-L),
\label{eq:phid}
\end{equation}
where the dependence of the fluxes and the oscillation probability on neutrino kinematics (energy, momentum) is implicit. The argument $t-L$ is the propagation delay, with the neutrinos taken to travel at essentially the speed of light, as is standard in oscillation experiments. We take Eq.~\eqref{eq:phid} as the definition of the time-independent oscillation probability $P_{\alpha\beta}(L)$, the quantity that relates the flux at the source to the flux at the detector at fixed energy and flavor. 

On the other hand, as we discuss in some detail in Appendix~\ref{app:t-integration},
\begin{equation}
\Phi_{\beta}^D(t) \propto \sum_\alpha \left(\int_{-\infty}^{\infty}~dt' \widetilde{P}_{\alpha\beta}(L,t-t')\Phi^S_{\alpha}(t')\right).
\label{eq:integral}
\end{equation}
The neutrino wave packets are sharply peaked at $t-t'=L/v_i$ where the different $v_i$ are very close to one another and to the speed of light. With this in mind, assuming $\Phi^S_{\alpha}(t)$ is relatively smooth, we can safely approximate 
\begin{equation}
\int_{-\infty}^{\infty}~dt' \widetilde{P}_{\alpha\beta}(L,t-t')\Phi^S_{\alpha}(t') \simeq
\Phi^S_{\alpha}(t-L)\int_{-\infty}^{\infty}~dt'' \widetilde{P}_{\alpha\beta}(L,t''),
\end{equation}
where, in the second expression, we changed the integration variable to $t''=t-t'$. Comparing this with the definition of $P_{\alpha\beta}(L)$ in Eq.~\eqref{eq:phid}, we identify
\begin{equation}
    P_{\alpha\beta}(L) \propto \int_{-\infty}^{\infty} dt''\,\widetilde{P}_{\alpha\beta}(L,t'')\,.
\end{equation}
That is, $P_{\alpha\beta}(L)$ is proportional to the time-averaged value of $\widetilde{P}_{\alpha\beta}$, which we denote
\begin{equation}
    \mathcal{P}_{\alpha\beta}(L) = \frac{1}{{\mathcal N}_T} \int_{-\infty}^{\infty} dt\; \widetilde{P}_{\alpha\beta}(L,t)
    =\frac{1}{\mathcal{N}_T}\int_{-\infty}^{\infty} dt \int_{-\infty}^{\infty} dx\, 
    e^{-\frac{(x-L)^2}{2\delta L^2}}\;|A_{\alpha\beta}(x,t)|^2\,,
    \label{eq:time_avg}
\end{equation}
with $1/\mathcal{N}_T$ a dimensionful normalization constant. Both the $dx$ and $dt$ integrals are Gaussian and can be performed analytically.

The quantity $\mathcal{P}_{\alpha\beta}(L)$ is the (time-averaged) probability that the neutrino is detected in the detection region \emph{and} is identified as flavor $\beta$. What an oscillation experiment reports, however, is conditioned on a neutrino having been detected at all. Given that a neutrino produced as $\nu_\alpha$ is observed in the detection region, the probability that it is observed as $\nu_\beta$ is  
\begin{equation}
    P_{\alpha\beta}(L) = \frac{\mathcal{P}_{\alpha\beta}(L)}{\sum_{\beta'} \mathcal{P}_{\alpha\beta'}(L)}\,.
    \label{eq:normalizing_prob}
\end{equation}
The denominator is the total probability of detecting a neutrino, independent of its flavor. Given Eq.~\eqref{eq:normalizing_prob}, it is easy to check that $\sum_{\beta} P_{\alpha\beta}(L) = 1$ for all $\alpha$ and $L$ and that $P_{\alpha\beta}(L)$ does not depend on $\mathcal{N}_T$ and other flavor-independent prefactors. 

Carrying out the Gaussian integrals in Eq.~\eqref{eq:time_avg}, we find
\begin{align} 
\label{eq:overlap_exact}
    \begin{split}
        \mathcal{P}_{\alpha \beta}(L)&= \frac{1}{\mathcal{N}_T}\sum_{i,j} U_{\alpha i}^*\, U_{\beta i}\, U_{\alpha j}\, U_{\beta j}^*\,\mathcal{N}_{ij}\,\\
        &\quad \times\exp\!\left(
        -\frac{L^2}{2\delta L^2 +  (L_{\text{coh}}^{ij})^2}
        - \frac{2\pi i\, L}{L_{\text{osc}}^{ij}}
        \frac{(L_{\text{coh}}^{ij})^2}{2\delta L^2 + (L_{\text{coh}}^{ij})^2}
        - c_{ij}
        \right) ,
    \end{split}
\end{align}
where we introduce the coherence and oscillation lengths
\begin{equation} 
\label{eq:Lcoh_Losc_def}
    L_{\text{coh}}^{ij}
    = 2 \sqrt{\frac{v_i^2 + v_j^2}{(v_i - v_j)^2}}\; \sigma_x ,
    \qquad
    L_{\text{osc}}^{ij}
    = \frac{2\pi}{\Delta E_{ij}}
    \left(
    \frac{v_i + v_j}{v_i^2 + v_j^2}
    - \frac{\Delta P_{ij}}{\Delta E_{ij}}
    \right)^{-1} ,
\end{equation}
along with the normalization factor
\begin{equation}
    \mathcal{N}_{ij}
    = \frac{2\sqrt{2}}{|v_i - v_j|}\frac{\sigma_x}{\sqrt{2 \delta L^2
    + (L_{\text{coh}}^{ij})^2}} ,
    \label{eq:def_Nij}
\end{equation}
and a baseline-independent constant
\begin{equation} 
\label{eq:cij_def}
    c_{ij}
    = \frac{
    2\left[
    (\Delta E_{ij} - v_i \Delta P_{ij})^2
    + (\Delta E_{ij} - v_j \Delta P_{ij})^2
    \right]
    \delta L^2 \sigma_x^2
    + 4\Delta E_{ij}^2\, \sigma_x^4}{
    (v_i - v_j)^2 [2 \delta L^2 + (L_{\text{coh}}^{ij})^2] 
    }\,.
\end{equation}

The three terms inside the exponential in Eq.~\eqref{eq:overlap_exact} play distinct roles. The first term, quadratic in $L$, controls the damping of the interference between different mass eigenstates driven by the spatial separation of the wave packets, which grows as the neutrinos propagate. The second term, linear in $L$, contains the oscillation phase through $L_{\text{osc}}^{ij}$. The third term is independent of the baseline and captures decoherence effects tied to the wave-packet and detector widths. 

In the regime relevant to oscillation experiments, $\delta L/L_{\text{coh}}^{ij}$ is negligible.
We will therefore safely take $\delta L / L_{\text{coh}}^{ij} \to 0$ henceforth. 
In this limit, the first and second terms reduce to the standard wave-packet decoherence damping and oscillation phase, respectively~\cite{Giunti:1991ca}. 
With all this in mind,  
\begin{align} 
\label{eq:Ptilde_simplified}
   \mathcal{P}_{\alpha \beta}(L)
    = \frac{1}{\mathcal{N}_T} \sum_{i,j}
    \mathcal{N}_{ij}\,U_{\alpha i}^*\, U_{\beta i}\, U_{\alpha j}\, U_{\beta j}^*\;
    \exp\!\left(-\frac{L^2}{  (L_{\text{coh}}^{ij})^2} -\frac{2\pi i\, L}{L_{\text{osc}}^{ij}} - c_{ij} \right).
\end{align}
Equation~\eqref{eq:Ptilde_simplified} already has the familiar form of an oscillation probability.
Here, we are interested in the phase factor $2\pi L/L_{\text{osc}}^{ij}$. According to Eq.~\eqref{eq:Lcoh_Losc_def}, $L_{\text{osc}}^{ij}$ is set by the mean momenta $P_i$ and energies $E_i(P_i)$ of the mass eigenstates. These depend on the kinematics of neutrino production, as we discuss in the next subsection.


\subsection{Production kinematics and the next-to-leading-order phase}
\label{sec:kinematics}

The oscillation length computed in the last subsection is a function of the mean energy $E_i(P_i)$ and momentum $P_i$ of each mass eigenstate. 
To obtain the NLO correction to the phase, these quantities cannot simply be related by following arbitrary equal-energy or equal-momentum prescriptions. 
Instead, they must be fixed by the kinematics of the production process. 
We carry this out in some detail here, concentrating on reactor antineutrinos.

Reactor antineutrinos are produced in nuclear beta decays, which are three-body decays. In principle, one could determine the mean energy and momentum of each neutrino mass eigenstate by averaging over the full three-body phase space. This would not, however, lead to a simple, closed-form expression, and the result would still have to be expanded to identify the dominant contributions to the oscillation phase. Instead, here we use an effective two-body description where we treat the final-state lepton and the daughter nucleus as a single object with invariant mass $m_{\rm eff}$, which varies event by event. We further assume that $m_{\rm eff}$ is the same for all the neutrino mass eigenstates. This simplification captures the leading dependence on the neutrino mass and allows us to identify the relative size of higher-order corrections that might vary event by event. We will argue that these are suppressed by ratios of the neutrino energy to the parent-nucleus mass. This approach can be found in the literature in a variety of circumstances.  For example, it is used in constraining the neutrino masses from the decays of taus into multiple pions and a neutrino~\cite{ALEPH:1997jrw, Perego:2002im}, where the hadronic system is similarly treated as a single particle. We therefore write the beta decay process as $N_A \rightarrow N_B + e^- + \bar{\nu}_e$ and treat the $N_B+e^-$ system as an effective recoiling particle of mass $m_{\rm eff}$. Taking the parent nucleus $N_A$ to decay at rest, the energy and momentum of the mass eigenstate $\nu_i$ are
\begin{equation}
    E_i(P_i) = \frac{m_A^2 + m_i^2 - m_{\rm eff}^2}{2m_A}\,,
    \qquad
    P_i = \frac{\sqrt{(m_A^2 + m_i^2 - m_{\rm eff}^2)^2 - 4m_A^2 m_i^2}}{2m_A} \,,
    \label{eq:2body_kinematics}
\end{equation}
where $m_A$ is the mass of the parent nucleus. As noted earlier, we assume that $m_\text{eff}^2$ does not depend on the neutrino mass index $i$.

With $E_i(P_i)$, $P_i$, and $v_i=P_i/E_i(P_i)$ fixed by the production kinematics, we can expand the oscillation phase in powers of the neutrino masses. This requires specifying a common energy or momentum scale around which the expansions are performed. At LO, expanding around any $E_i(P_i)$, $P_i$, or another common reference scale yields the same result. Beyond LO, however, different choices lead to different expressions. Once experimental observables are properly identified, the quantitative impact of the NLO correction should not, of course, depend on the choice of the reference scale. Nonetheless, as is often the case, some choices lead to more transparent final answers. Here, the most convenient choice turns out to be $E_0$, the energy the neutrino would have if it were massless. For beta decays,
\begin{equation}
E_0 = \frac{m_A^2 - m_{\rm eff}^2}{2m_A} .
\label{eq:E0_def}
\end{equation}
$E_0$ is the quantity most directly related to the neutrino energy reconstructed in a reactor experiment, as we discuss later.

Performing a fixed order expansion up to $\mathcal{O}(m_i^4/E_0^4)$, we find
\begin{align}
        \label{eq:DEij}\Delta E_{ij} &= \frac{\Delta m^2_{ij}}{2m_A},\\
        \label{eq:DPij}\Delta P_{ij} &= -\frac{\Delta m^2_{ij}}{2E_0} + \frac{\Delta m^2_{ij}}{2m_A} -\frac{(m^4_i-m^4_j)}{8E_0^3} + \frac{(m^4_i-m^4_j)}{4E_0^2m_A }  + \mathcal{O}\left(\frac{m_i^6}{E_0^5}\right), \\
    v_i & = \frac{P_i}{E_i} = 1-\frac{m_i^2}{2E_0^2} - \frac{m_i^4}{8E_0^4}\left(1-\frac{4E_0}{m_A}\right) + \mathcal{O}\left(\frac{m_i^6}{E_0^6}\right),
\end{align}

Substituting into the expression for $L_{\rm osc}^{ij}$ in Eq.~\eqref{eq:Lcoh_Losc_def} allows us to expand the oscillation phase systematically in powers of the neutrino masses. Keeping terms through $\mathcal{O}(m_i^4/E_0^3)$, we find\footnote{We are abusing notation here, for the sake of clarity. In reality, we are expanding $2\pi/(E_0L^{ij}_{\rm osc})$ in powers of $m_i^2/E_0^2$ up to $\mathcal{O}(m_i^4/E_0^4)$.}
\begin{equation}
    \frac{2\pi}{L_{\rm osc}^{ij}}
    = \frac{\Delta m^2_{ij}}{2E_0}
    + \frac{m_i^4 - m_j^4}{16 E_0^3} \biggl(1 + \frac{m_{\rm eff}^2}{m_A^2} \biggr)
    + \mathcal{O}\left( \frac{m_i^6}{E_0^5}\right).
    \label{eq:phase_NLO_general}
\end{equation}
The LO result for the oscillation length is the familiar one. The NLO correction depends on the production process through the ratio $m_{\rm eff}^2/m_A^2$. We can rewrite it in a more convenient form by noting that
$m_{\rm eff}^2 + m_A^2 = 2m_A(m_A - E_0)$, so
\begin{equation}
    \frac{2\pi}{L_{\rm osc}^{ij}}
    = \frac{\Delta m^2_{ij}}{2E_0}
    + \frac{m_i^4 - m_j^4}{8\,E_0^3}
    \left(1 - \frac{E_0}{m_A}\right)
    + \mathcal{O}\left( \frac{m_i^6}{E_0^5}\right).
    \label{eq:phase_NLO}
\end{equation}
For beta decays of interest to commercial nuclear reactors, $E_0$ is around a few to several MeV while $m_A$ are nuclear masses (several tens to a few hundred GeV), so $E_0/m_A \lesssim 10^{-4}$ and can be safely neglected. Hence, while the oscillation length depends on the details of the nuclear decay, it is the same, for a fixed neutrino energy, at LO and virtually the same, up to permille corrections, at NLO. 

Physically, the $m_i^4$ structure of the new term has an intuitive origin: it is the next order in the relativistic expansion of the neutrino energy,
\begin{equation}
E_i(P_i) = \sqrt{P_i^2+m_i^2} \simeq P_i + \frac{m_i^2}{2P_i} - \frac{m_i^4}{8P_i^3}.
\end{equation}
The $m_i^2$ piece yields the standard $\Delta m_{ij}^2/2E$ phase, and the $m_i^4$ piece contributes to the new correction. A proper derivation, however, requires the wave-packet treatment, careful consideration of the neutrino-production kinematics, and the choice of $E_0$ as the expansion parameter, developed above. See Appendix~\ref{app:NLOplanewaves} for detailed discussions.

On the other hand, 
\begin{align}
    \begin{split}
        L_{\rm coh} &= \frac{4\sqrt{2} E_0^2 \sigma_x}{\Delta m_{ij}^2}
        \left[1 - \frac{m_i^2 + m_j^2}{2 E_{0}^2} + \frac{m_i^2 + m_j^2}{m_AE_{0}}\right.\\
        &\quad \left.- \frac{(m_i^2 - m_j^2)^2}{32 E_{0}^4} - \frac{m_i^2 m_j^2}{m_A E_{0}^3} + \frac{m_i^4 + 5 m_i^2 m_j^2 + m_j^4}{4 m_A^2E_{0}^2} + \mathcal{O}\left(\frac{m_i^6}{E_0^6}\right)\right].
    \end{split}
\end{align}
The leading term is the usual form of the coherence length. Decoherence effects have never been observed in terrestrial experiments and the leading term is, for the foreseeable future, sufficient for all phenomenological applications. 

The factor $\mathcal{N}_{ij}$ in Eq.~\eqref{eq:Ptilde_simplified}, defined in Eq.~\eqref{eq:def_Nij}, is flavor-independent at LO and its contribution disappears once the probabilities are properly normalized. 
At next-to-leading order, it acquires a mass-dependent correction. 
However, these corrections are of $\mathcal{O}(\Delta m^2_{ij}/E_0^2)$, and, most importantly, they modify the oscillation amplitudes rather than the phase. Therefore, their effects are negligible in oscillation experiments.

Similarly, we will ignore the effect of $c_{ij}$ on oscillation probabilities. In a little more detail, in the limit $\delta L \rightarrow 0$, Eq.~\eqref{eq:cij_def} becomes
\begin{equation} 
\label{eq:cijnodeltaL}
    c_{ij} = \frac{(\Delta E_{ij})^2 \sigma_x^2}{\left( v_i^2 + v_j^2 \right)} ,
\end{equation}
and, at LO,
\begin{equation} 
\label{eq:cijwithDeltaP}
    c_{ij} \simeq \frac{(\Delta m_{ij}^2)^2 \sigma_x^2}{8m_A^2} .
\end{equation}
Superficially, this result is surprising as it is inversely proportional to the mass of the parent nucleus instead of the neutrino energy. This indicates that, even at LO, $c_{ij}$ depends on the details of the production process: reactor neutrinos produced in beta decays of different parent nuclei have different $c_{ij}$, while neutrinos with different energies but the same parents have, at LO, the same $c_{ij}$. While intriguing, results consistent with ours have been found in the literature \cite{Giunti:1991ca, Giunti:1997wq}. We refer to these publications for detailed discussions. 

Putting everything together, in the limit where decoherence effects are negligible, the oscillation probability inferred by a reactor neutrino experiment, including the NLO $m_i^4L/E_0^3$ corrections, is
\begin{align} 
\label{eq:JUNO_prob}
   P_{\alpha \beta}(L)
    = \sum_{i,j}
    U_{\alpha i}^*\, U_{\beta i}\, U_{\alpha j}\, U_{\beta j}^*\;
    \exp\!\left[-i\left( \frac{\Delta m^2_{ij}L}{2E_0}
    + \frac{(m_{i}^4 - m_{j}^4) L}{8\,E_{0}^3} \right)\right]
    .
\end{align}
%


\subsection{Experimental signatures}
\label{subsec:experimental_signatures}

In order to illustrate the impact of the NLO corrections to the oscillation phase on neutrino oscillations, we examine the two-flavor transition probability (neutrino masses $m_1,m_2$, mass-squared difference $\Delta m^2=m_2^2-m_1^2$, mixing angle $\theta$ ($U_{\alpha1}=U_{\beta2}=\cos\theta$, $U_{\alpha2}=-U_{\beta1}=\sin\theta$)).
For $\alpha\neq\beta$,
\begin{equation}
    P_{\alpha\beta}
    = \sin^22\theta \sin^2\!\left[\frac{\Delta m^2\, L}{4 E_0} \left(1 + \frac{m_1^2 + m_2^2}{4 E_0^2}\right)\right].
    \label{eq:Posc_LO_2flavor}
\end{equation}

The NLO correction to the oscillation phase deviates from the traditional $L/E_0$ scaling of the LO contribution, and the oscillation probability depends on $L$ and $E_0$ separately. A measurement of the oscillation phase that distinguishes the LO and NLO contributions is sensitive to both $\Delta m^2$ and $m_1^2+m_2^2$ and would, therefore, allow independent measurements of both $m_1$ and $m_2$. 

As an aside, we comment that, in vacuum, it is not possible to distinguish the sign of $\Delta m^2$ even if one includes the NLO correction to the oscillation phase. This lack of sensitivity is an ``all orders'' result and is easy to understand. For two flavors, there are two ways to define the parameter space such that all physically distinguishable masses and mixing angles are uniquely represented. One is to restrict $\theta\in[0,\pi/4]$ and allow both positive and negative $\Delta m^2$, and the other is to allow $\theta\in[0,\pi/2]$ and restrict the sign of $\Delta m^2$, e.g., $m_2^2>m_1^2$. Since the vacuum oscillation probability cannot distinguish $\theta$ from $\pi/2-\theta$, it must also not be able to distinguish $\Delta m^2$ from $-\Delta m^2$. For more details, see, for example, Ref.~\cite{deGouvea:2008nm}. As is well known, this degeneracy is not preserved in matter.

\begin{figure}[t]
    \centering
    \includegraphics[width=0.7\linewidth]{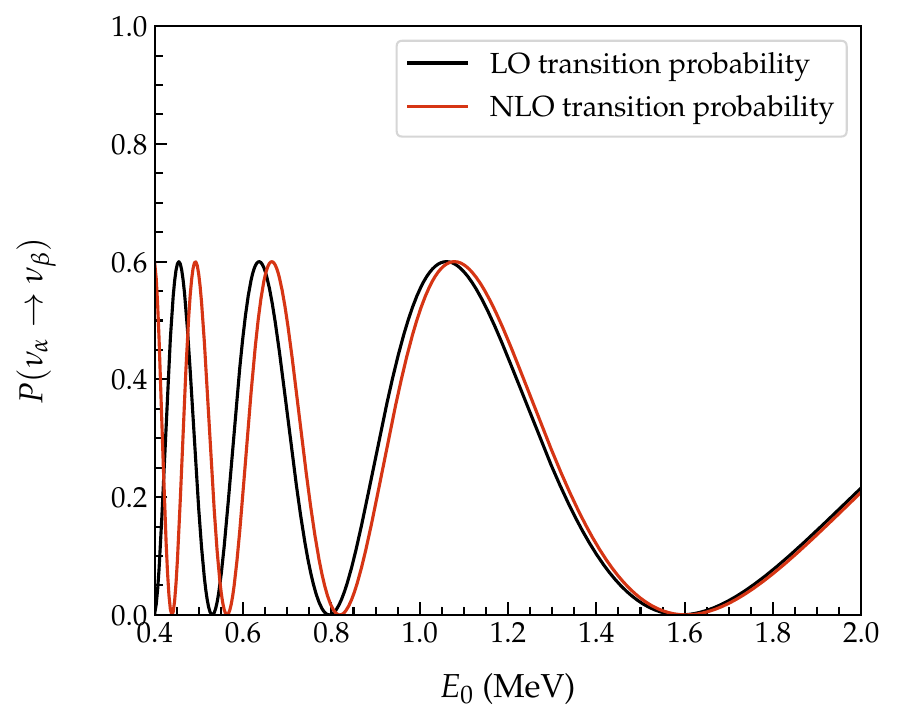}
    \caption{Two-flavor transition probability as a function of neutrino energy $E_0$ at a fixed baseline $L$ of 52.5~km, using illustrative oscillation parameters with $\sin^2 (2\theta) = 0.6$, $\Delta m^2=7.5 \times 10^{-5}$~eV$^2$, and $m_1 = 200$~keV. The two curves coincide at high $E_0$ and progressively dephase toward low $E_0$, where the fractional next-to-leading-order shift $m_1^2/(2 E_0^2)$ grows.}
    \label{fig:NLO_LoverE}
\end{figure}

Figure~\ref{fig:NLO_LoverE} depicts the two-flavor transition probability as a function of $E_0$ at a fixed baseline $L=52.5$~km, using illustrative oscillation parameters chosen to make the NLO effect visible: $\sin^2 2\theta = 0.6$, $\Delta m^2=7.5\times 10^{-5}$~eV$^2$ and $m_1 = 200$~keV.

The NLO correction translates into an energy-dependent shift of the oscillation frequency and does not impact the oscillation amplitude. This distinguishes it from other effects often discussed in the context of neutrino oscillations, including wave-packet decoherence and detector energy smearing, both of which suppress the amplitude of the oscillation while leaving its frequency intact, and matter effects, which modify both the oscillation length and the amplitude in an energy-dependent way. 

Figure~\ref{fig:NLO_LoverE} also reveals that NLO effects are more pronounced at low energies. This of course makes sense because the NLO correction is proportional to $m_i^2/E_0^2$. Sensitivity to NLO effects and potentially to the absolute neutrino mass scale requires excellent energy resolution at the lowest neutrino energies. We turn to JUNO in the next section.


\section{JUNO Sensitivity to the Absolute Neutrino Mass}
\label{sec:JUNO_study}

The Jiangmen Underground Neutrino Observatory (JUNO) provides a unique opportunity to place a constraint on the neutrino masses based on neutrino oscillations. JUNO is a 20~kton (fiducial) liquid scintillator detector located at a baseline of 52.5~km from the Taishan and Yangjiang reactor complexes~\cite{JUNO:2022mxj}. It detects reactor $\bar{\nu}_e$ via inverse beta decay, reconstructing the neutrino energy from the positron signal. With its first 59.1 days of data, JUNO has already delivered the world's most precise measurements of $\sin^2\theta_{12}$ and $\Delta m^2_{21}$~\cite{JUNO:2025gmd}, demonstrating the detector's performance~\cite{JUNO:2025fpc}. Its combination of long baseline, superb energy resolution, and large expected statistics makes it an ideal candidate for this search. 

JUNO measures the survival probability of electron anti-neutrinos. Including NLO corrections, in vacuum,
\begin{align} 
    \label{eq:Pee_exact}
        P_{\bar{\nu}_e \to \bar{\nu}_e} &= 1 - \sin^2 2\theta_{12} \cos^4 \theta_{13} \sin^2 \!\left[\frac{\Delta m^2_{21}\, L}{4 E_0} \left(1 + \frac{m_1^2 + m_2^2}{4 E_0^2}\right)\right] \nonumber\\
        & \hspace{0.7cm}- \sin^2 2\theta_{13} \left\{ \cos^2 \theta_{12} \sin^2\left[\frac{\Delta m_{31}^2\, L}{4 E_0} \left(1 + \frac{m_1^2 + m_3^2}{4 E_0^2}\right)\right]\right. \\ 
        & \hspace{2.9cm}+ \left. \sin^2 \theta_{12} \sin^2 \!\left[\frac{\Delta m_{32}^2\, L}{4 E_0} \left(1 + \frac{m_2^2 + m_3^2}{4 E_0^2}\right)\right] \right\} \nonumber
\end{align}
In our simulations, we ignore matter effects. While matter effects impact JUNO's precise determination of the oscillation parameters~\cite{Khan:2019doq}, we expect that they will not meaningfully hinder our ability to constrain the absolute neutrino mass scale. First, matter effects impact, in a correlated way, the oscillation length and the amplitude; the NLO effects discussed here only impact the oscillation length. Second, assuming the matter potential $A=\sqrt{2}G_Fn_e$ ($G_F$ is the Fermi constant and $n_e$ is the average electron number density of the medium) is much smaller than $\Delta m^2_{21}/2E_0$, the relative matter-induced shift to the oscillation phase is proportional to $E_0A/\Delta m^2_{21}$ while the relative NLO effect is proportional to $(m_i^2+m_j^2)/E_0^2$. The energy dependences of the two effects are quite different: one grows linearly with energy while the other falls off like $1/E_0^2$. We repeated the analysis discussed in the next subsection including matter effects, assuming the matter potential is known precisely. The bound on $m_1$ changes by less than $1\%$. 


\subsection{JUNO analysis}

JUNO detects reactor antineutrinos via inverse beta decay (IBD), $\bar{\nu}_e + p \to n + e^+$~\cite{JUNO:2022mxj}. The positron produced in the interaction deposits its energy in the detector and is measured by an array of photomultiplier tubes. The prompt energy deposited in the detector, $E_p$, is directly related to the incoming neutrino energy in the following way. 
First, the neutrino energy is related to the total positron energy through $E_0 \simeq E_{e} + \Delta$, where $E_0$ is the energy of a massless neutrino, $E_{e}$ is the total energy of the positron, and $\Delta = m_n - m_p$. The recoil energy of the neutron, to leading order in $1/m_n$, is $(\vec{p}_\nu - \vec{p}_e)^2/2m_n$, which is of order a few to tens of keV. We neglect it in our nominal analysis. Including the recoil correction consistently shifts the mass bound by less than 0.3\%.
The prompt energy, $E_p$, is the kinetic energy of the positron plus that of the annihilation photons, assuming all the energy is deposited in the detector. This can be approximated as $E_p = E_0 - \Delta + m_e$.

The expected reconstructed energy spectrum of IBD events as a function of the visible energy, $E_\text{vis}$, is~\cite{JUNO:2022mxj, Forero:2021lax}
\begin{equation} 
\label{eq:JUNO_spectrum}
    S(E_\text{vis}) = N_p\epsilon\int_{T_\text{DAQ}} dt
    \int_{1.8~\text{MeV}}^{12~\text{MeV}} dE_0\;
    \Phi(E_0, t)\,  \sigma(E_0)\,  R(E_0, E_\text{vis}) \,,
\end{equation}
where $N_p$ is the number of free protons in the target, $\epsilon$ is the IBD selection efficiency, $T_\text{DAQ}$ is the data-taking period, and $R(E_0, E_\text{vis})$ is the detector energy response function. The energy response function maps the anti-neutrino energy to the visible energy via the prompt energy $E_p$ and accounts for the energy transfer in the IBD reaction, the non-linearity of the detector, and the energy resolution of the detector (for more detail, see Ref.~\cite{JUNO:2022mxj}). JUNO's energy resolution is taken to be $\sigma_E / E_\text{vis} \simeq 3\%/\sqrt{E_\text{vis}[\text{MeV}]}$~\cite{JUNO:2024fdc,JUNO:2022mxj}. With a detection rate of roughly 57 events per day, a combined IBD selection efficiency of 82.2\%, six years of operation, and a reactor duty cycle factor of 11/12 to account for refueling, we expect approximately $9.5 \times 10^4$ events.

The standard IBD cross section was computed following Refs.~\cite{Ricciardi:2024cit, Vogel:1999zy} and is
\begin{equation}
\label{eq:IBDsigma}
    \frac{d\sigma}{d\cos\theta} = \frac{\sigma_0}{2}
    \left[(f^2 + 3g^2) + (f^2 - g^2)\, v_e \cos\theta \right] E_e\, p_e \,,
\end{equation}
where $v_e=p_e/E_e$, $\sigma_0 = G_F^2 \cos^2\theta_C (1 + \Delta_\text{inner}^R)/\pi$, $f$ and $g$ are hadronic form factors, $G_F = 1.166 \times 10^{-5}\ \text{GeV}^{-2}$ is the Fermi constant, $\theta_C = 13.02^\circ$ is the Cabibbo angle, and $\Delta_\text{inner}^R = 0.024$ is the energy-independent radiative correction that arises from short wavelength virtual interactions at the quark level. Equation~\eqref{eq:IBDsigma} assumes the neutrino is massless and neglects nuclear recoil. Neglecting the recoil, we can approximate $f = 1$ and $g = 1.26$. Still neglecting the neutron recoil but accounting for a massive neutrino, we find
\begin{equation} 
\label{eq:IBDsigmamassiveneutrinos}
    \frac{d\sigma}{d\cos\theta} = \frac{\sigma_0}{2}
    \left[(f^2 + 3g^2) + (f^2 - g^2)\, v_e v_{\bar{\nu}_e} \cos\theta \right] E_e\, p_e\, \frac{1}{v_{\bar{\nu}_e}}\, ,
\end{equation}
where $v_{\bar{\nu}_e}$ is the velocity of the electron antineutrino.

The oscillated reactor antineutrino flux, summed over reactor cores $r$, is~\cite{JUNO:2022mxj}
\begin{equation} \label{eq:reactor_flux}
    \Phi(E_0) \simeq \sum_r \frac{P(\bar{\nu}_e \to \bar{\nu}_e; E_0, L_r)}{4\pi L_r^2}\,
    \frac{W_r}{\sum_j f_{jr}\, e_j}
    \sum_j f_{jr}\, s_j(E_0) \,.
\end{equation}
Here, $W_r(t)$ is the thermal power of reactor $r$, $f_{jr}(t)$ is the fission fraction for isotope $j$ (${}^{235}$U, ${}^{238}$U, ${}^{239}$Pu, ${}^{241}$Pu), $e_j$ is the mean energy released per fission, and $s_j(E_0)$ is the antineutrino energy spectrum per fission $j$ parametrized by the Huber--Mueller model~\cite{Mueller:2011nm, Huber:2011wv} as
\begin{equation}
    s_j(E_0) = \exp\!\left(\sum_{k=1}^{6} \alpha_{jk}\, E_0^{k-1}\right),
\end{equation}
and the values used for $\alpha_{jk}$ can be found in Table VI from Refs.~\cite{Mueller:2011nm, Huber:2011wv}. This spectrum also assumes massless neutrinos. If we account for the neutrino mass, we find the beta decay spectrum is modified by an overall factor of $v_{\bar{\nu}_e}$.  
Thus, when the cross section and beta decay spectrum are combined, their dependence on the neutrino mass cancels, leaving the reconstructed IBD energy spectrum unchanged. For more detail, see Appendix~\ref{app:xsecandfluxcancellation}.

We then construct the chi-squared function with systematics, following Refs.~\cite{JUNO:2022mxj, Forero:2021lax}. The event spectrum is binned into 325 bins of 20~keV in visible energy between 0.94 and 7.44~MeV~\cite{JUNO:2022mxj}. For systematics, we include a correlated detector uncertainty, a bin-to-bin uncorrelated shape uncertainty, and the detector's non-linear energy response. The shape uncertainty has an energy dependence following Fig.~6 of Ref.~\cite{JUNO:2022mxj}.

\begin{figure}[t]
    \includegraphics[width=0.7\linewidth]{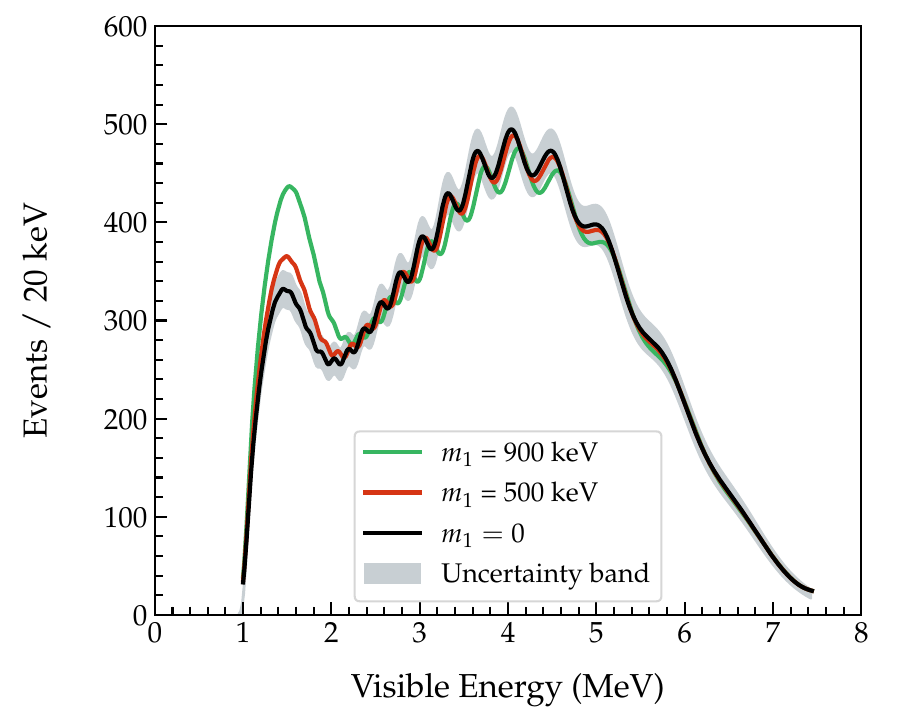}
    \caption{Expected number of events per 20 keV bin at JUNO as a function of the visible energy $E_{\text{vis}}$ for 6 years of running time, for $m_1$  values of 0, 500~keV, and 900~keV. The grey band indicates the $1\sigma$ statistical uncertainty.}
    \label{fig:masssensitivitybands}
\end{figure}

Figure~\ref{fig:masssensitivitybands} depicts the expected number of events per 20~keV bin in JUNO over six years as a function of the visible energy, assuming the normal neutrino-mass ordering ($m_1<m_2<m_3$), for different values of the lightest neutrino mass: $m_1 = 0, 500$~keV, and $900$~keV.
We assume oscillation parameters from NuFit v6.0~\cite{Esteban:2024eli}.
For small values of $m_1$, we accurately reproduce JUNO's expected event spectrum in Ref.~\cite{JUNO:2022mxj}. The figure indicates that JUNO should be sensitive to nonzero values of $m_1$ that are larger than a few hundred keV.

\begin{figure}
    \centering
    \includegraphics[width=0.8\linewidth]{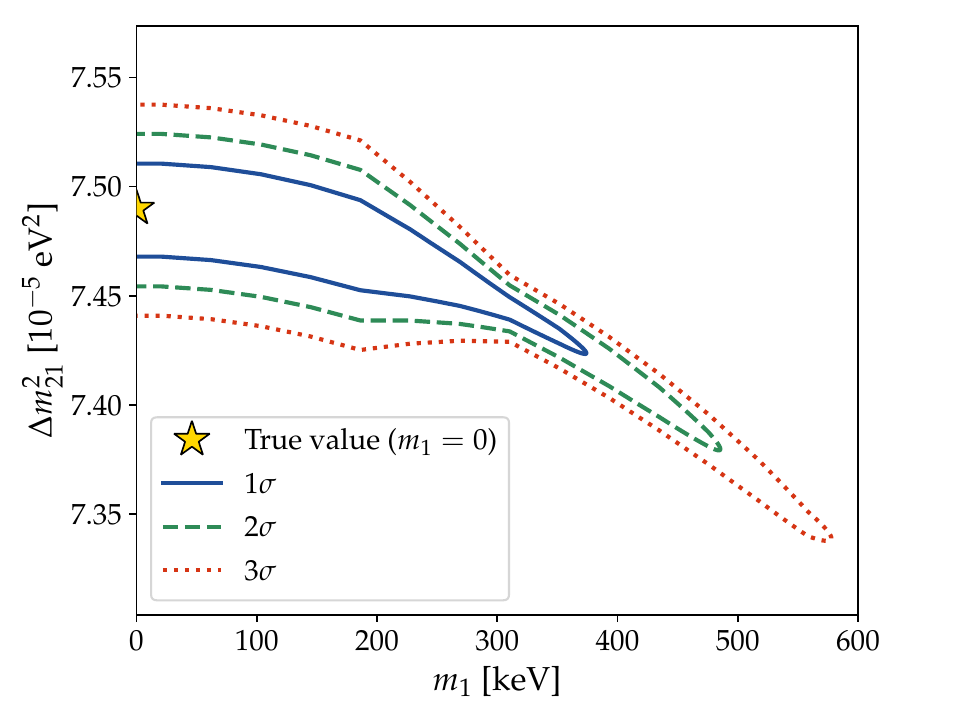}
    \caption{The $1\sigma$, $2\sigma$, and $3\sigma$ contours in the $m_1$--$\Delta m_{21}^2$ plane correspond to $\Delta\chi^2 = 2.30$, $6.18$, and $11.83$, respectively. The star indicates the true values, $m_1 = 0$ and $\Delta m_{21}^2 = 7.49 \times 10^{-5}\,\mathrm{eV}^2$, used to generate the pseudo-data.}
    \label{fig:dm21vsm1contour}
\end{figure}

In order to estimate JUNO's sensitivity to the absolute mass scale, we perform a $\chi^2$ fit to simulated data consistent with the NLO model, see Eq.~\eqref{eq:Pee_exact}, with $m_1 = 0$ and the oscillation parameters given by NuFit v6.0~\cite{Esteban:2024eli} for the normal ordering. These are $\Delta m_{31}^2 = 2.513 \times 10^{-3}~\text{eV}^2$, $\Delta m_{21}^2 = 7.49 \times 10^{-5}~\text{eV}^2$, $\sin^2{\theta_{12}} = 0.308$, and $\sin^2{\theta_{13}} = 0.02215$.  The fit model uses the NLO probability with $m_1^2$ as an additional free parameter on top of the standard oscillation parameters $\Delta m^2_{31}$, $\Delta m^2_{21}$, $\sin^2\theta_{12}$, and $\sin^2\theta_{13}$ with the mass ordering fixed. We then construct the profiled $\Delta\chi^2$ as a function of $m_1^2$ by minimizing over all other oscillation parameters for each value of $m_1$. The $\Delta\chi^2(m_1^2) = 1~(4)$ contour corresponds to the $1\sigma$ ($2\sigma$) sensitivity. We find that JUNO should be able to constrain $m_1<305$~keV at $1\sigma$ and $m_1<434$~keV at $2\sigma$, without imposing external priors on the oscillation parameters. The sensitivity is limited by a parameter degeneracy: at any single energy, the NLO correction is indistinguishable from a rescaling $\Delta m^2_{21} \to \Delta m^2_{21}\,(1 + m_1^2/2E_0^2)$, so the fit can absorb much of the signal into a shift of $\Delta m^2_{21}$ and the mixing angles. Only the $1/E_0^2$ energy dependence of the NLO correction breaks this degeneracy. Figure~\ref{fig:dm21vsm1contour} shows the $1\sigma$, $2\sigma$, and $3\sigma$ contours in the $m_1$--$\Delta m_{21}^2$ plane. If all oscillation parameters are instead held fixed at their true values, the bound improves to $m_1 < 162$~keV at $1\sigma$.

Imposing Gaussian priors on the oscillation parameters from NuFit~v6.0~\cite{Esteban:2024eli}, the latest global fit without JUNO data, does not make a significant difference to the results. The bounds become $m_1 < \rm 294$~keV at $1\sigma$ and $m_1 < \rm 419$~keV at $2\sigma$.


\subsection{Origin of the absolute mass sensitivity}
\label{sec:sector_decomposition}

\begin{figure}[t]
    \centering
    \includegraphics[width=0.7\linewidth]{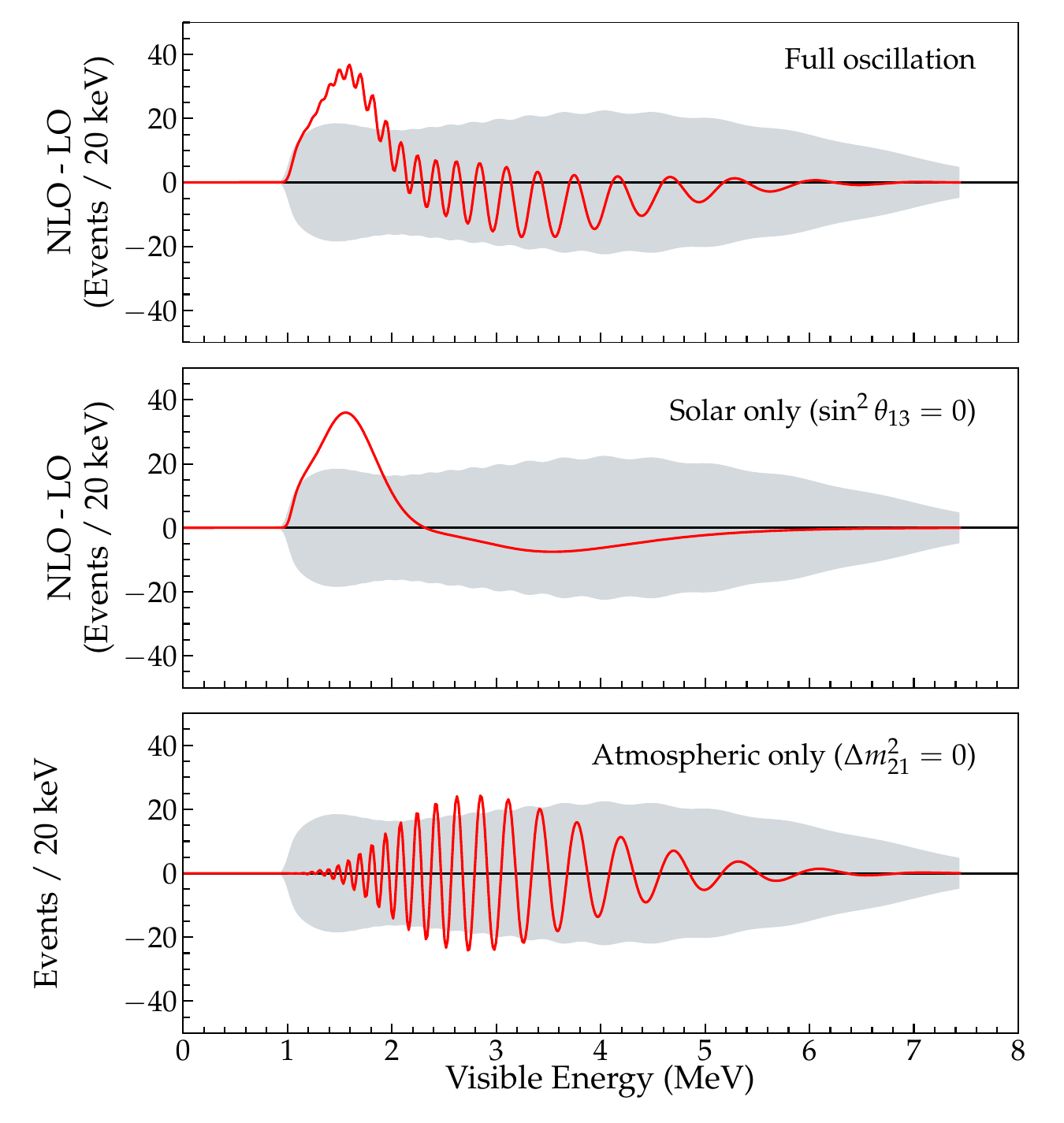}
    \caption{Difference between the NLO and LO event rates (per 20 keV energy bin) for 6 years of run time, for $m_1 = 500$~keV and $L = 52.5$~km, decomposed by oscillation channel. The grey bands show the statistical uncertainty for the NLO event spectrum, for visual guidance.}
    \label{fig:sector_decomposition}
\end{figure}

Having established JUNO's sensitivity, it is natural to ask which oscillation channel drives it. The NLO term in Eq.~\eqref{eq:Pee_exact} modifies the phase of every channel proportionally to $m_1^2/E_0^2$, so in principle the sensitivity comes from both the solar ($\Delta m^2_{21}$) and atmospheric ($\Delta m^2_{31}$, $\Delta m^2_{32}$) contributions to $P(\bar\nu_e \to \bar\nu_e)$. 
To identify which channel contributes the most to the sensitivity, we compute the IBD event spectrum keeping only one channel at a time. The ``solar-only'' spectrum is obtained by setting $\sin^2\theta_{13} = 0$. This eliminates the contributions from the two atmospheric terms in Eq.~\eqref{eq:Pee_exact} while keeping the slow $\Delta m^2_{21}$ oscillation with amplitude $\sin^2 2\theta_{12}$. The ``atmospheric-only'' spectrum is obtained by setting $\Delta m_{21}^2 = 0$, which forces $\Delta m_{32}^2 = \Delta m_{31}^2$. This eliminates the solar envelope and merges the two atmospheric terms into a single fast oscillation with amplitude $\sin^2 2\theta_{13}$. Figure~\ref{fig:sector_decomposition} depicts the difference between the NLO and LO event rates per energy bin for $m_1=500$~keV assuming the full oscillation probability, the ``solar-only'' case, and the ``atmospheric-only'' case. 
The grey bands show the statistical uncertainty for the NLO event spectrum, for visual guidance.

The NLO effects have comparable magnitudes in the two oscillation channels, both about 30 events per bin. In both channels, the NLO correction shifts the phases by the same fraction $m_1^2/2E_0^2$, but the absolute shift is proportional to the phase itself, so the atmospheric shift is larger by a factor $\Delta m_{31}^2/\Delta m_{21}^2 \simeq 34$. This more than offsets the factor of ten by which the solar amplitude ($\simeq$~0.81) exceeds the atmospheric one ($\simeq$~0.085), so the two residuals end up comparable once the rapid atmospheric oscillation is smeared by the detector resolution.
The solar-channel residual reproduces essentially all of the slow envelope visible in the full result, while the atmospheric residual is rapidly oscillating about zero.
In terms of the sensitivity, the two channels are close, but the solar one leads: repeating the analysis with the NLO correction applied to only the atmospheric channel (only the solar channel) yields a $1\sigma$ bound of 394~keV (347~keV), compared to 305~keV for the full three-flavor analysis.

This edge of the solar channel is set by two features specific to JUNO's setup. 
The first is that the solar dip ($L/E_0 \sim 16.5$~km/MeV at its first minimum) lands directly in JUNO's high-statistics IBD energy band of $1.8$--$8$~MeV, so the $m_1$-induced shift of the dip overlaps with the bins that have the largest event counts. The second is that JUNO's energy resolution $\sigma_E / E_0 \simeq 3\%/\sqrt{E_0[\text{MeV}]}$ produces a smearing of $\sim 60$~keV at $E_0 = 4$~MeV, which is smaller than but non-negligible compared with a single atmospheric oscillation period ($\sim 300$~keV at this energy), partially averaging the fast atmospheric pattern, while the solar dip is much broader than $\sigma_E$ and survives smearing intact.


\section{Summary of Existing Absolute Mass Probes}
\label{sec:mass_limits}

The most straightforward constraints on the absolute mass scale come from the kinematics of decay and capture processes. These constraints translate into bounds on weighted sums of the neutrino masses, $m_{\nu_\alpha}^2 = \sum_{i} |U_{\alpha i}|^2 m_i^2$, $\alpha=e,\mu,\tau$. In beta decays, a nonzero neutrino mass impacts the electron energy spectrum near its endpoint. The KATRIN experiment probes the spectral endpoint and places an upper limit $m_{\nu_e} < 0.45$~eV at 90\% C.L.~\cite{KATRIN:2024cdt}. Looking ahead, the Project~8 experiment aims to reach sensitivities of $m_{\nu_e} < 0.04$~eV using cyclotron radiation emission spectroscopy to measure the beta decays of atomic tritium~\cite{Project8:2022hun, Project8:2025hjp}. The ECHo~\cite{ECHo:2025ook} and HOLMES~\cite{Alpert:2025tqq} experiments use the endpoint of the spectrum of electron capture of $^{163}$Ho to limit the neutrino mass. ECHo places an upper bound $m_{\nu_e} < 15$~eV at 90\% C.L., while HOLMES sets $m_{\nu_e} < 27$~eV at 90\% C.L. 

In two-body decays, the energy of the outgoing particles is completely fixed by the kinematics of the initial state and the masses of the particles in the decay. The mass of the neutrino directly impacts the energy of the daughter particle produced along with it. Although less competitive, the study of pion decay at rest yields $m_{\nu_\mu} < 0.17$~MeV at 90\% C.L.~\cite{Assamagan:1995wb}, while tau decays with many pions in the final state yield $m_{\nu_\tau} < 18.2$~MeV~\cite{ALEPH:1997jrw} (ALEPH collaboration) and $m_{\nu_\tau} < 48$~MeV~\cite{Perego:2002im} (DELPHI collaboration), both at 95\% C.L.

Another independent probe of the neutrino mass is provided by the time-of-flight delay of neutrinos from astrophysical sources. A massive neutrino with mass $m$ and energy $E$ traveling a distance $L$ arrives with a delay $\Delta t = (L/2)\times (m/E)^2$ relative to a massless neutrino. Since different supernova neutrinos have different energies, a measurement of their distribution as a function of arrival time can be used to constrain the neutrino mass.   
This method was applied to the neutrinos detected from the supernova SN1987A~\cite{Loredo:2001rx, Pagliaroli:2010ik}, yielding an upper bound of $m < 5.8$~eV at 95\% C.L. When information from neutrino oscillation experiments is included, this bound applies to $m_1,m_2,m_3$ individually.

Massive neutrinos also leave distinctive imprints on the evolution of the universe~\cite{Lesgourgues:2013sjj}. In the early universe, neutrinos are relativistic and contribute to the radiation energy density, influencing the expansion rate during big bang nucleosynthesis and recombination. As the universe expands, they transition to non-relativistic speeds and begin contributing to the matter density instead. In addition, the large thermal velocities of neutrinos cause them to free-stream out of overdense regions, suppressing the growth of structure below a characteristic scale. Observations of large-scale structure formation, the cosmic microwave background (CMB), and baryon acoustic oscillations (BAO) are sensitive to these effects and are used to constrain both the sum of the neutrino masses, $\sum m_i$, and the number of relativistic species, $N_{\mathrm{eff}}$.
 
By combining CMB anisotropy data from Planck~\cite{Planck:2019nip, Carron:2022eyg} and the Atacama Cosmology Telescope (ACT)~\cite{ACT:2023kun, ACT:2023dou} with BAO measurements from the Dark Energy Spectroscopic Instrument (DESI)~\cite{DESI:2013agm, DESI:2022xcl, DESI:2016igz}, current analyses~\cite{Elbers:2025vlz} yield upper bounds on the sum of neutrino masses that depend on the assumed cosmological model. 
Within the $\Lambda$CDM model, the upper limit depends on the prior used for the neutrino mass. They find $\sum m_i < 0.053$~eV at 95\% C.L. when the neutrino masses are constrained to the physical region. In the standard $\Lambda$CDM model, dark energy is a cosmological constant. Allowing it to evolve with redshift, as in the $w_0 w_a$CDM model, introduces additional freedom that is partially degenerate with the neutrino mass, relaxing the constraint considerably to $\sum m_i < 0.163$~eV (95\% C.L.). These bounds can also be significantly affected if neutrinos participate in interactions beyond those of the Standard Model~\cite{Beacom:2004yd, Bellomo:2016xhl,  Escudero:2020ped, Esteban:2021ozz, FrancoAbellan:2021hdb}.

If neutrinos are Majorana particles, an additional constraint on their mass comes from searches for neutrinoless double-beta decay ($0\nu\beta\beta$). 
In this process, two neutrons in a nucleus decay simultaneously into two protons and two electrons with no emitted neutrinos, violating lepton number by two units. The decay rate depends on the effective Majorana mass, $m_{\beta\beta} = |\sum_i U_{ei}^2 m_i|$. The non-observation of this process by the KamLAND-Zen experiment sets a lower limit on the half-life of $T_{1/2}^{0\nu} > 3.8 \times 10^{26}$~yr at 90\% CL, which translates into an upper bound on the effective Majorana mass in the range $m_{\beta\beta} < 0.028$--$0.122$~eV, where the spread reflects uncertainties in the nuclear matrix element calculations~\cite{KamLAND-Zen:2024eml}. Similar upper bounds have been reported by the LEGEND collaboration, $m_{\beta\beta} < 0.075$--$0.200$~eV \cite{LEGEND:2025jwu}.

While all of these approaches aim to constrain the neutrino mass, each one probes a different combination of the masses and relies on distinct physical assumptions. As fundamental properties of the neutrino sector, such as the mass ordering, the Dirac or Majorana nature, and the potential existence of interactions beyond the Standard Model, remain open questions, it is important to constrain the underlying neutrino mass scale using independent approaches. 
Agreement across methods would strengthen the robustness of the result, while tensions could point to interesting routes to be pursued. 

The JUNO sensitivity estimated here is quantitatively much weaker than most existing bounds. Instead, its significance is that it arises directly from the propagation phase and probes different combinations of $m_i^2 + m_j^2$ through the corresponding oscillation channels. It therefore represents a conceptually distinct absolute mass probe.


\section{Conclusions}
\label{sec:conclusions}

We computed the NLO corrections to the neutrino oscillation phase, which are of order $m_i^4/E_0^3$,
using a wave-packet framework and carefully addressing the conceptual subtleties that arise when going beyond the LO approximation. These include carefully considering the kinematics of the neutrino production process. Along the way, we also computed the NLO corrections to different decoherence parameters. These corrections to the decoherence parameters are negligible for the phenomenological applications considered here.

The NLO correction to the oscillation length carries a different dependence on the neutrino masses relative to the LO contribution, and reveals that oscillation experiments are, in principle, sensitive to both the neutrino mass-squared differences and neutrino mass-squared sums, and hence the individual values of the neutrino masses. 

Given the ultrarelativistic nature of neutrinos, NLO effects are very small, suppressed relative to the LO contribution by $m_i^2/E_0^2\sim 10^{-14}~(m_i^2/{10^{-2}~\rm eV^2})(1~{\rm MeV}/E_0)^2$. Nonetheless, we used the NLO expression to estimate JUNO's sensitivity to the absolute neutrino masses. We find that JUNO data are sensitive to $m_1 \gtrsim 305$~keV at $1\sigma$. While this is not competitive with existing bounds from beta decay, electron capture, or cosmology, it probes a distinct combination of the neutrino masses through a fundamentally different physical mechanism and provides complementary information to other approaches.

The NLO correction derived here may be of practical use for other oscillation systems, including the oscillations of hypothetical, quasi-degenerate heavy neutrinos, such as those that may be responsible for the baryon asymmetry of the universe (see, for example, Ref.~\cite{Davidson:2008bu} for an overview), and the oscillation of very-low-energy neutrinos. For example, tritium beta decay has a $Q$-value under 20~keV and the antineutrinos from the heavily scrutinized end-point of the decay spectrum have energies around 1~eV. Needless to say, to date, these have never been directly observed. 


\acknowledgments

G.F.S.A.\ and P.A.N.M.\ are partially supported by Fermi Forward Discovery Group, LLC under Contract No. 89243024CSC000002 with the U.S. Department of Energy, Office of Science, Office of High Energy Physics. G.F.S.A.\ and A.L.C.d.G.\ are partially supported by the U.S. Department of Energy, Office of Science, Office of High Energy Physics under Grant No.~DE-SC0010143. The work of J.K.\ and S.W.L.\ was supported in part by the National Science Foundation under Grant No.\ PHY-2544442. A.L.C.d.G.\ and S.W.L.\ also acknowledge the Center for Theoretical Underground Physics and Related Areas (CETUP), The Institute for Underground Science at Sanford Underground Research Facility (SURF), and the South Dakota Science and Technology Authority for hospitality and financial support while some of this work was carried out.

\appendix


\section{Plane-Wave Approach to NLO Corrections}
\label{app:NLOplanewaves}

In the main text we derived the NLO corrections to neutrino oscillations within a wave-packet framework.
In this appendix we show that the naive plane-wave approach, which yields the correct result at LO, fails at NLO.
The shortcomings of the plane-wave approach to neutrino oscillations are well known~\cite{Kayser:1981ye, Giunti:1991ca, Beuthe:2001rc, Akhmedov:2009rb, Giunti:2003ax, Akhmedov:2019iyt} and, at NLO, these shortcomings become apparent.

In the plane-wave description of neutrino oscillations, the time-evolved flavor state is given by
\begin{equation}
    | \nu_\alpha (x, t) \rangle = \sum_i U_{\alpha i}^* e^{i (p_i x - E_i t)} | \nu_i \rangle ,
\end{equation}
where $p_i$ and $E_i$ are the momentum and energy of the $i$-th mass eigenstate. The oscillation probability is then
\begin{equation} 
\label{eq:planewaveprob}
    P_{\alpha \beta} = |\langle \nu_\beta | \nu_\alpha(x,t) \rangle |^2 = \sum_{i,j} U_{\alpha i}^* U_{\beta i} U_{\alpha j} U_{\beta j}^* e^{i (\Delta p_{ij} x - \Delta E_{ij} t)} .
\end{equation}
At this stage, one of two assumptions is typically made: either the neutrinos have equal momenta ($p_i = p_j$) or equal energies ($E_i = E_j$) for all $i,j$.
Both assumptions are inherently inconsistent since both the average energy and momentum of the different mass eigenstates differ at order $\mathcal{O}(m^2/E^2)$; see, for example, Eq.~(\ref{eq:DEij}) and Eq.~\eqref{eq:DPij}. 
Proceeding nonetheless, under the equal-momentum assumption, the phase in Eq.~\eqref{eq:planewaveprob} is
\begin{equation} 
\label{eq:planewaveprobdeltaE}
    -i\Delta E_{ij} t = -i\left( E_i - E_j \right) t = - i\left[\frac{\Delta m_{ij}^2}{2p} - \frac{m_i^4 - m_j^4}{8p^3} + {\cal O}\left(\frac{m_i^6}{p^5}\right)\right] t ,
\end{equation}
while under the equal-energy assumption, taking $x = L$, one finds
\begin{equation} 
\label{eq:planewaveprobdeltap}
    i\Delta p_{ij} L = i\left( p_i - p_j \right) L = -i\left[ \frac{\Delta m_{ij}^2}{2E} + \frac{m_i^4 - m_j^4}{8E^3} + {\cal O}\left(\frac{m_i^6}{E^5}\right) \right] L .
\end{equation}
These two expressions cannot be compared directly, since Eq.~\eqref{eq:planewaveprobdeltaE} depends on the propagation time $t$, while Eq.~\eqref{eq:planewaveprobdeltap} depends on the baseline $L$. Relating $t$ to $L$ requires information about the propagation velocity of each mass eigenstate, but plane waves carry no well-defined group velocity, making the replacement $t \to L$ ambiguous, regardless of the order of the expansion.
Nevertheless, at LO, the standard approximation is to take $t = L$. At LO this leads to an unambiguous result since deviations from $v_i=v_j=1$ are $\mathcal{O}(m^2/E^2)$.
At NLO, these types of corrections are precisely what we wish to retain, so the prescription matters. We consider three representative choices below.

\begin{enumerate}

    \item $v_i = c$ for all $i$. Setting $t = L$, as at LO, assumes all mass eigenstates travel at the speed of light. This contradicts the NLO expansion from the outset, and substituting $t = L$ into Eq.~\eqref{eq:planewaveprobdeltaE} yields the opposite sign on the NLO term relative to Eq.~\eqref{eq:planewaveprobdeltap}.

   \item $t_i = L/v_i$ for each mass eigenstate. Under the equal-momentum assumption, $v_i = p/E_i$, giving $t_i = E_i L/p$. Different mass eigenstates are then assigned different propagation times, and the result depends on when the time-to-baseline replacement is performed relative to the mass expansion.

    Substituting $t_i = E_i L/p$ before expanding the energies yields
    \begin{equation}
        -i\frac{(E_i^2 - E_j^2)\,L}{p} = -i\frac{\Delta m_{ij}^2\,L}{p} ,
    \end{equation}
    which is twice the LO phase and contains no NLO correction.
    Expanding the energies first, then replacing the time and carrying out the expansion of the velocities, yields, ignoring terms ${\cal O}(m^6_i/p^5)$,
    \begin{align}
    \label{eq:option2after}
        \begin{split}
            -i\Delta E_{ij}t &= -i\left[ \left(p + \frac{m_{i}^2}{2p}-\frac{m_i^4}{8p^3}\right)t  -\left(p + \frac{m_{j}^2}{2p}-\frac{m_j^4}{8p^3}\right)t  \right]\\
            &=-i\left[\left(\frac{m_{i}^2}{2p}-\frac{m_i^4}{8p^3}\right)\frac{L}{v_i}-\left(\frac{m_{j}^2}{2p}-\frac{m_j^4}{8p^3}\right)\frac{L}{v_j}\right]\\
            &= -i\left[\frac{\Delta m_{ij}^2}{2p} + \frac{m_i^4 - m_j^4}{8p^3}\right]L .
        \end{split}
    \end{align}
    This appears to reproduce the correct LO and NLO structure, but two major problems remain. First, the two orderings disagree: using a common time for both mass eigenstates in the first line of Eq.~\eqref{eq:option2after} cancels the momentum-dependent term, which is precisely why this result differs from the one obtained before. Second, the denominator involves $p$ rather than $E$, and replacing $p \to E$ is not self-consistent at this stage, as it introduces corrections to the LO term of the same order as the NLO term itself. This prescription therefore fails as well.

    \item $t = 2L/(v_i + v_j)$ (average velocity). One may instead use a single propagation time based on the average velocity of the two mass eigenstates. The oscillation phase in this case is
    \begin{equation}
        -\left[\frac{\Delta m_{ij}^2\,L}{2p} - \frac{(\Delta m_{ij}^2)^3 L}{32 p^5} \right],
    \end{equation}
    where the correction is of order $\mathcal{O}(m^6/E^5)$, beyond the order we are considering. 
    Therefore, this prescription misses the $\mathcal{O}(m^4/E^3)$ correction entirely.
\end{enumerate}

Among all cases considered, only Eq.~\eqref{eq:option2after}, applied after expanding the energies, reproduces the correct LO and NLO structure, yet all the steps involved in deriving it are wrong for the reasons noted above. 

In summary, the plane-wave approach faces two distinct problems at NLO: imposing equal momenta (or equal energies) is itself inconsistent at this order, and the prescription for relating $t$ to $L$ is ambiguous, with different choices yielding qualitatively different phases. Most prescriptions give the correct LO phase, showing that at LO many approximations lead to the same result.
A wave-packet treatment, combined with the kinematics of the production and detection processes, is therefore the framework for describing neutrino oscillations at NLO, as we showed in the main text.


\section{On the Neutrino Flux at the Detector}
\label{app:t-integration}

In the text, we discussed how, in the context of a neutrino oscillation experiment, the probability $P_{\alpha\beta}(L,t,t')$ that a neutrino produced as a $\nu_{\alpha}$ at the origin of space at time $t'$ is measured as a $\nu_{\beta}$ at a position $L$ at a time $t$ is related to the time-independent probability $P_{\alpha\beta}(L)$ that a neutrino produced as a $\nu_{\alpha}$ will be detected as a $\nu_{\beta}$ after it traverses a distance $L$. Here we expand on this issue. Throughout, $\alpha,\beta=e,\mu,\tau$. We assume space is one-dimensional and we ignore the neutrino kinematics and the detailed physics of the measurement process. 

Experiments measure (time integrals of) neutrino fluxes of the different flavors inside the detector, so we are interested in computing the flux $\phi_{\beta}^D(t)$ of $\nu_{\beta}$ at the detector site given a flux $\phi_{\alpha}^S(t)$ of $\nu_{\alpha}$ at the source. 

$\phi_{\beta}^D(t)$ is proportional to $P_{\alpha\beta}(L,t,t') \phi_{\alpha}^S(t')$, summed over all production times: 
\begin{equation}
\phi_{\beta}^D(t) = C_{\alpha}\int_{-\infty}^{+\infty} dt'~P_{\alpha\beta}(L,t,t') \phi_{\alpha}^S(t'),
\label{eq:phid_app}
\end{equation}
where $C_{\alpha}$ is a proportionality constant. For convenience, we assume that the source produces only one neutrino species (in this case, $\nu_{\alpha}$). In order to compute $\phi^D_{\beta}$ for sources that emit several neutrino flavors, one needs to sum the right-hand side of Eq.~\eqref{eq:phid_app} over all values of $\alpha$. We return to this momentarily. Note that while Eq.~\eqref{eq:phid_app} allows for $t'$ values larger than $t$ to contribute to $\phi_{\beta}^D(t)$, it does not imply that causality is violated: $P_{\alpha\beta}(L,t,t')$ controls whether it is possible for neutrinos to arrive at the detector before they were produced at the source.  

The constant $C_{\alpha}$ is chosen such that the total number $N_{\rm tot}^D$ of neutrinos at the detector is the same as the total number $N_{\rm tot}^S$ of neutrinos emitted at the source.  
Using Eq.~\eqref{eq:phid_app},
\begin{eqnarray}
N_{\rm tot}^D & = & \sum_{\beta}\int_{-\infty}^{+\infty} dt~\Phi_{\beta}^{D}(t), \nonumber \\
& = & \sum_{\beta}\int_{-\infty}^{+\infty} dt \left(C_{\alpha} \int_{-\infty}^{+\infty} dt'~P_{\alpha\beta}(L,t,t') \phi_{\alpha}^S(t')\right), \nonumber \\
& = & \int_{-\infty}^{+\infty} dt'~ \phi_{\alpha}^S(t')
\left(C_{\alpha}\sum_{\beta}\int_{-\infty}^{+\infty}dt~P_{
\alpha\beta}(L,t,t')\right).
\end{eqnarray}
$P_{\alpha\beta}(L,t,t')$ depends only on $t''=t-t'$. We take advantage of this to write
\begin{equation}
N_{\rm tot}^D = 
\left(C_{\alpha}\sum_{\beta}\int_{-\infty}^{+\infty}dt''~P_{
\alpha\beta}(L,t'')\right)N_{\rm tot}^{S},
\end{equation}
where the total number of neutrinos emitted at the source is
\begin{equation}
N_{\rm tot}^{S} = \int_{-\infty}^{+\infty} dt~\Phi_{\alpha}^{S}(t).
\label{eq:N_app}
\end{equation} 
Hence, $N_{\rm tot}^S =  N_{\rm tot}^D$ implies
\begin{equation}
C_{\alpha} = \frac{1}{\sum_{\beta}\int_{-\infty}^{+\infty}dt~P_{
\alpha\beta}(L,t,t')}.
\label{eq:Calpha}
\end{equation}

If different flavors are produced at the source, each with flux $\phi^S_{\alpha}(t)$,
\begin{equation}
\phi_{\beta}^D(t) = \sum_{\alpha}C_{\alpha}\int_{-\infty}^{+\infty} dt'~P_{\alpha\beta}(L,t,t') \phi_{\alpha}^S(t'),
\end{equation}
where $C_{\alpha}$ is still given by Eq.~\eqref{eq:Calpha}. This implies not only that the total number of neutrinos emitted at the source is the same as the number of neutrinos that go through the detector, but it also implies that the number of neutrinos is conserved for each individual initial flavor: If a total $N^S_\alpha$ of $\nu_{\alpha}$ is produced at the source, the total number of neutrinos of all flavors at the detector associated with the initial-state flavor $\nu_{\alpha}$ is also $N^S_{\alpha}$.    

To illuminate our result, it is useful to consider two examples. If the flux at the source is constant, 
\begin{eqnarray}
\phi_{\beta}^D &=& \sum_{\alpha} \phi_{\alpha}^S\frac{\int_{-\infty}^{+\infty} dt'~P_{\alpha\beta}(L,t,t')}{\sum_{\beta'}\int_{-\infty}^{+\infty}dt~P_{
\alpha\beta'}(L,t,t')}, \\
&=& \sum_{\alpha} \phi_{\alpha}^S P_{\alpha\beta}(L),
\end{eqnarray}
where, keeping in mind that $P_{\alpha\beta}(L,t,t')=P_{\alpha\beta}(L,t-t')$ and defining $t-t'=t''$,
\begin{equation}
P_{\alpha\beta}(L) \equiv \frac{\int_{-\infty}^{+\infty} dt''~P_{\alpha\beta}(L,t'')}{\sum_{\beta'}\int_{-\infty}^{+\infty}dt''~P_{
\alpha\beta'}(L,t'')}.
\end{equation}
This is the result used in the text: The time-independent $P_{\alpha\beta}(L)$ is the average of $P_{\alpha\beta}(L,t-t')$ over all propagation times $t-t'$. As expected, the time-independent probability is properly normalized, $\sum_{\beta}P_{\alpha\beta}(L) = 1$, $\forall\alpha$.

Finally, let us consider the scenario where there are no flavor oscillations so $P_{\alpha\beta}(L,t-t')=\delta_{\alpha\beta}P(L,t-t')$ and where a sudden burst of neutrinos $\nu_{\alpha}$ is emitted from the source at $t=0$: $\Phi^S_{\alpha}(t)=N_0\delta(t)$. In this case,
\begin{equation}
\Phi_{\alpha}^D(L,t) = N_0 \frac{P(L,t)}{\int_{-\infty}^{+\infty}dt''~P(L,t'')}.
\end{equation}
The flux at the detector inherits the shape of $P(L,t-t')$ and is normalized such that the total number of neutrinos that can ever go through the detector -- in this case, $N_0$ -- is preserved. 


\section{IBD Cross Section and Beta Decay Spectrum for Massive Neutrinos} 
\label{app:xsecandfluxcancellation}

According to Eq.~\eqref{eq:IBDsigmamassiveneutrinos}, nonzero antineutrino masses modify the IBD cross section in two ways: The $(f^2 - g^2)v_e\cos\theta$ term acquires an extra factor of $v_{\bar\nu_e}$, and the overall cross section is rescaled by $1/v_{\bar\nu_e}$, related to the incident-flux normalization. We are interested in the total cross section. Upon integrating over $\cos\theta$, the angular-dependent term vanishes and we are left with 
\begin{equation} 
\label{eq:IBDNLO}
    \sigma_\text{IBD} (E_e) = \frac{\sigma_0}{v_{\bar{\nu}_e}} (f^2 + 3g^2)\, p_e E_e ~.
\end{equation}

The beta decay spectrum of the $b^\text{th}$ branch of the $j^\text{th}$ fission product follows from the phase space factor $p_e E_e p_{\bar{\nu}_e} E_{\bar{\nu}_e}$, where $p_e$ ($p_{\bar{\nu}_e}$) and $E_e$ ($E_{\bar{\nu}_e}$) are the momentum and energy of the electron (antineutrino); the remaining factors are independent of $E_{\bar{\nu}_e}$~\cite{Mueller:2011nm, Huber:2011wv}. Energy conservation fixes $E_{\bar{\nu}_e} = (m_A - m_\text{eff}) - E_e$, where $m_A$ is the mass of the parent nucleus and $m_\text{eff}$ is the effective mass of the system after the decay, so
\begin{equation}
\label{eq:phasespacecalc}
    p_e E_e p_{\bar{\nu}_e} E_{\bar{\nu}_e}
    = p_e E_e \frac{p_{\bar{\nu}_e}}{E_{\bar{\nu}_e}} E_{\bar{\nu}_e}^2
    = p_e E_e v_{\bar{\nu}_e} \left( (m_A - m_\text{eff}) - E_e \right)^2 .
\end{equation}

For a massless antineutrino, $v_{\bar{\nu}_e} = 1$, and the end-point energy of the branch is $E_{0,j}^b = m_A - m_\text{eff}$ at leading order in $E_{\bar{\nu}_e}/m_A$; Eq.~\eqref{eq:phasespacecalc} then reduces to the familiar spectrum $S_j^b(E_e) \propto p_e E_e (E_{0,j}^b -
E_e)^2$~\cite{Mueller:2011nm, Huber:2011wv}. For a massive neutrino, the only change is the extra overall factor of $v_{\bar{\nu}_e}$.

The flux in Eq.~\eqref{eq:reactor_flux} is a linear combination of the spectra of the individual beta decay branches, and each branch carries the
same overall factor of $v_{\bar{\nu}_e}$. The factor therefore propagates directly to the flux: $\Phi(E_0) = v_{\bar{\nu}_e} \Phi_{m_\nu = 0}(E_0)$,
where the subscript denotes the massless-neutrino expression. Combined with the cross section in Eq.~\eqref{eq:IBDNLO},
\begin{align}
    \sigma_\text{IBD}(E_0) \Phi(E_0) &= \frac{1}{v_{\bar{\nu}_e}}
    \sigma_{\text{IBD}, m_\nu = 0}(E_0) \times v_{\bar{\nu}_e}
    \Phi_{m_\nu = 0}(E_0) \nonumber \\
    &= \sigma_{\text{IBD}, m_\nu = 0}(E_0) \times \Phi_{m_\nu = 0}(E_0) ~.
\end{align}
Hence the factors cancel, leaving the expected IBD event spectrum unchanged.


\bibliographystyle{JHEP}
\bibliography{biblio.bib}

\end{document}